%% file: QBFF_QAI.tex
\documentclass[conference,a4paper]{IEEEtran}
\IEEEoverridecommandlockouts

\usepackage[utf8]{inputenc}
\usepackage[T1]{fontenc}
\usepackage{cite}
\usepackage{amsmath,amssymb,amsfonts}
\usepackage{graphicx}
\usepackage{textcomp}
\usepackage{xcolor}
\usepackage{booktabs}
\usepackage{colortbl}
\usepackage{multirow}
\usepackage{array}
\usepackage{enumitem}
\usepackage{comment}
\usepackage{braket}
\usepackage{stfloats}
\usepackage[font=small]{caption}
\usepackage{url}
\usepackage[hidelinks]{hyperref}
\usepackage{tabularx}
\usepackage{tikz}
\usetikzlibrary{arrows.meta,positioning,fit,calc}
\DeclareUnicodeCharacter{00B1}{$\pm$}
\DeclareUnicodeCharacter{2019}{'}
\DeclareUnicodeCharacter{2013}{--}
\DeclareUnicodeCharacter{2014}{---}
\DeclareUnicodeCharacter{2192}{$\rightarrow$}

\renewcommand{\emph}[1]{\textit{#1}}

\usepackage[misc]{ifsym}

\def\BibTeX{{\rm B\kern-.05em{\sc i\kern-.025em b}\kern-.08em
    T\kern-.1667em\lower.7ex\hbox{E}\kern-.125emX}}

\newcommand{\sd}[1]{\,{\scriptsize\textcolor{black!50}{$\pm$#1}}}
\definecolor{relcol}{gray}{0.92}
\newcolumntype{R}{>{\columncolor{relcol}}c}

\colorlet{goodcol}{green!22}
\colorlet{badcol}{red!17}
\newcommand{\cg}{\cellcolor{goodcol}}
\newcommand{\cb}{\cellcolor{badcol}}

\begin{document}

\title{QBFF: Deep Quantum Bottleneck for Automatic Speaker Verification Anti-Spoofing}

% \author{\IEEEauthorblockN{Anonymous Authors}}

\author{%
\IEEEauthorblockN{Anton Firc\IEEEauthorrefmark{4},
Nouhaila Innan\IEEEauthorrefmark{2}\IEEEauthorrefmark{3},
Vojt\v{e}ch Stan\v{e}k\IEEEauthorrefmark{4},
Kamil Malinka\IEEEauthorrefmark{4},
Martin Pere\v{s}\'{i}ni\IEEEauthorrefmark{4}\\
Zbyn\v{e}k Li\v{c}ka\IEEEauthorrefmark{4},
Walid El Maouaki\IEEEauthorrefmark{2},
Alberto Marchisio\IEEEauthorrefmark{2}\IEEEauthorrefmark{3}\IEEEauthorrefmark{5}$^*$\thanks{$^*$This work has been carried out while at NYUAD.},
Muhammad Shafique\IEEEauthorrefmark{2}\IEEEauthorrefmark{3}}
\IEEEauthorblockA{\IEEEauthorrefmark{4}Security@FIT, Brno University of Technology, Brno, Czech Republic}
\IEEEauthorblockA{\IEEEauthorrefmark{2}eBRAIN Lab, Division of Engineering, New York University Abu Dhabi (NYUAD), Abu Dhabi, UAE}
\IEEEauthorblockA{\IEEEauthorrefmark{3}Center for Quantum and Topological Systems (CQTS), NYUAD Research Institute, NYUAD, Abu Dhabi, UAE}
\IEEEauthorblockA{\IEEEauthorrefmark{5}SDU Microelectronics, Institute of Mechanical and Electrical Engineering, University of Southern Denmark, Odense, Denmark}
\IEEEauthorblockA{\parbox{0.95\textwidth}{\centering\footnotesize
Email: \{ifirc,iperesini,malinka,istanek,ilicka\}@fit.vut.cz\\
\{nouhaila.innan,walid.el.maouaki,muhammad.shafique\}@nyu.edu, chisio@sdu.dk}}
}

\maketitle

\begin{abstract}
A central goal of quantum machine learning is to identify applications in which small quantum circuits can improve task performance while remaining executable on real hardware. We investigate deepfake (synthetic) speech detection, a largely unexplored application in this area, where reliable confidence scores are as important as discrimination performance because deployment decisions are made at fixed thresholds and overconfident errors can be costly.
We introduce \textit{QBFF}, which routes a compressed speech representation through a small $8$-qubit variational circuit (VQC) used as a drop-in bottleneck, with a matched classical bottleneck as the control. The quantum circuit yields better-calibrated confidence in most matched comparisons, including out-of-distribution evaluation, though the downstream deployment-cost benefit is model-dependent. We then run the same circuit on a real IBM quantum processor. To our knowledge, this is the first anti-spoofing pipeline to run on a QPU; hardware execution demonstrates feasibility, with lower subset EER and $C_{llr}$ but higher mean actDCF than the matched classical reference. Together, these results show that hardware-executable quantum bottlenecks can improve score reliability, with gains that vary by evaluation setting.
\end{abstract}


%----------------------------------------
\section{Introduction}
\label{sec:intro}

Variational quantum circuits (VQCs) are increasingly studied as trainable function classes for near-term quantum machine learning (QML)~\cite{biamonte2017quantum}, but their role inside large-scale classical pipelines remains insufficiently understood. A key open question is whether shallow, hardware-compatible circuits can improve system-level behavior beyond standard predictive accuracy when used as constrained intermediate transformations. This question is especially relevant in decision systems where deployment risk depends on the quality of the output score. We study this setting through a security application that remains largely unexplored in QML: \emph{detecting deepfake speech}.

High-quality synthetic speech can now be generated from only a few seconds of a target speaker, enabling impersonation attacks against automatic speaker verification systems~\cite{deepfakesurvey_firc24}. Deepfake speech detection has therefore become a major audio-security problem~\cite{asvspoofdataset}. Unlike standard binary classification, anti-spoofing systems are evaluated through scores that are converted into decisions by a fixed threshold. Two models with similar ranking performance can therefore incur different deployment costs when their score distributions differ~\cite{wang24_asvspoof}. As a result, reliable detection depends on both class separation and the calibration and scale of the learned scores. We use this score-based setting to study whether a constrained quantum bottleneck can reshape the output behavior of a classical speech pipeline.

% \begin{figure}
%     \centering
% \includegraphics[width=1\linewidth]{img/Contribution.pdf}
%     \caption{Summary of QBFF’s key contributions. The blue boxes indicate the proposed components.}
%     \label{fig:contribution}
%     \vspace{-1em}
% \end{figure}

\input{img/abstract_fig}

Prior quantum anti-spoofing studies do not yet answer this question. One hybrid system uses custom data and reports accuracy~\cite{pandey2025hybridcqdeepfakeaudio}; the quantum-kernel method reports an $18.3\%$ equal error rate (EER; lower means better separation) on ASVspoof~5~\cite{alamins2025qksvm}; and the VQC transfer model reports $5.51\%$ EER on ASVspoof~2021~\cite{wang2023qtlwavlm}, where modern classical detectors reach roughly $2$--$3\%$~\cite{tak22_odyssey}. These studies also do not combine matched quantum--classical control, score-calibration analysis, and real-QPU execution. Our matched comparison makes the benefit concrete: in the strongest out-of-domain \texttt{AASIST} setting, the quantum bottleneck cuts EER by $28\%$ relative ($5.49\%\rightarrow3.97\%$) and fixed-threshold actual detection cost (actDCF; lower is better) by $12\%$ ($0.205\rightarrow0.181$) compared with the classical bottleneck.

We compare quantum and classical transformations of a compressed speech representation. A self-supervised front-end produces a pooled $1024$-dimensional utterance embedding~\cite{rohdin24_asvspoof}, which both heads project to $16$ bounded values. \textit{QBFF} maps these through an $8$-qubit VQC to eight Pauli-$Z$ expectation values; \textit{BFF} uses a classical mixer with matching input and output dimensions, nominal depth, and classifier interface. This comparison tests how the mixer choice changes detector scores.

The main effect is observed in score reliability rather than in discrimination alone. Across ASVspoof evaluation protocols~\cite{wang24_asvspoof}, \textit{QBFF} often matches the equal error rate (EER) of \textit{BFF} while reducing deployment-cost metrics, including actual detection cost (actDCF) and log-likelihood-ratio cost ($C_{llr}$), in several matched settings, including out-of-distribution evaluation. This behavior is consistent with the bounded range of quantum measurement expectations, which constrains the bottleneck output scale and may reduce overconfident posteriors before the final classifier. We observe these gains with a fixed circuit design, without circuit architecture or circuit-specific hyperparameter search.

% The same trained circuit is then executed on IBM Heron quantum processors. Under finite-shot inference with error mitigation, the reliability trend persists on a matched hardness-stratified subset: at $512$ shots, actDCF decreases from $0.227$ for the matched classical bottleneck to $0.144$ for \textit{QBFF}, and $C_{llr}$ decreases from $1.045$ to $0.312$. \textbf{To our knowledge, this is the first speech anti-spoofing pipeline evaluated on a real QPU.}

The same trained circuit is then executed on IBM Heron quantum processors. On a hardness-stratified subset, finite-shot inference yields lower mean EER and $C_{llr}$ than the classical reference, while actDCF varies across hardware runs. \textbf{To our knowledge, this is the first speech anti-spoofing pipeline evaluated on a real QPU.}

\noindent The contributions of this work are as follows (see \autoref{fig:contribution}):
\begin{itemize}[leftmargin=*]
\item We propose \textit{QBFF}, a hybrid quantum--classical anti-spoofing head that uses an $8$-qubit variational quantum circuit within a $1024\rightarrow16\rightarrow8$ feature bottleneck inside a self-supervised speech pipeline.\footnote{\url{https://github.com/Security-FIT/QBFF}.}
\item We introduce \textit{BFF}, a classical control matched to \textit{QBFF} in bottleneck width, depth, and classifier interface to isolate the quantum transformation.
\item We show that \textit{QBFF} can improve out-of-domain generalization over matched \textit{BFF}: on In-the-Wild with \texttt{AASIST}, it reduces EER by $28\%$ relative ($5.49\%\rightarrow3.97\%$) and actDCF by $12\%$ ($0.205\rightarrow0.181$).
\item We demonstrate \textit{QBFF} on IBM Heron-class QPUs: across five $512$-shot runs on a hardness-stratified ASVspoof~5 subset, it achieves lower mean EER and $C_{llr}$ than the classical reference, while actDCF varies substantially.
\end{itemize}

%----------------------------------------
\section{Background and Related Work}
\label{sec}

% This section introduces the speech anti-spoofing task, the score-based evaluation protocol, and the quantum speech studies most closely related to our work. The emphasis is on the distinction between discrimination and score reliability, as it motivates the proposed quantum bottleneck.

\subsection{Synthetic speech detection}
% Modern text-to-speech and voice-conversion systems can clone a target voice from a few seconds of audio. Attackers use the resulting synthetic, or \emph{deepfake}, speech to impersonate people and to defeat automatic speaker verification (ASV) systems, which decide whether a recording matches an enrolled speaker~\cite{speech-biometric-practicalattack_firc22}. A \emph{spoofing countermeasure}, also called an anti-spoofing system or spoof detector, is a separate binary classifier that guards the ASV system. It takes one audio recording, and labels it \emph{bona fide} (genuine human speech) or \emph{spoof} (synthetic or manipulated speech)~\cite{wang24_asvspoof}.

% The detector does not emit a hard label directly. For each recording, it produces a real-valued \emph{score}, and a fixed \emph{decision threshold} turns that score into an accept-or-reject decision: scores on one side of the threshold are read as bona fide, scores on the other side as spoof. The threshold is chosen before deployment based on the expected costs of the two error types. The quality of the scores and the placement of the threshold, therefore, matter as much as the underlying ranking of recordings~\cite{BRUMMER2006230}.

Synthetic speech can impersonate enrolled speakers and threaten automatic speaker verification (ASV)~\cite{speech-biometric-practicalattack_firc22}. A spoofing countermeasure classifies recordings as \emph{bona fide} or \emph{spoof}~\cite{wang24_asvspoof}. It produces a score that is compared with a decision threshold chosen for the deployment conditions and error costs. Detection therefore depends on both score quality and threshold selection~\cite{BRUMMER2006230}.

\subsection{Discrimination and score reliability}

\emph{Discrimination} describes how well scores separate bona fide from spoof recordings. \emph{Calibration} describes whether predicted probabilities agree with observed frequencies: among recordings assigned $90\%$ bona fide probability, approximately $90\%$ should be genuine. Good discrimination can coexist with poor calibration, leading to errors at a fixed deployment threshold. We therefore evaluate both.

\begin{figure*}[t]
    \centering
    \includegraphics[width=1\linewidth]{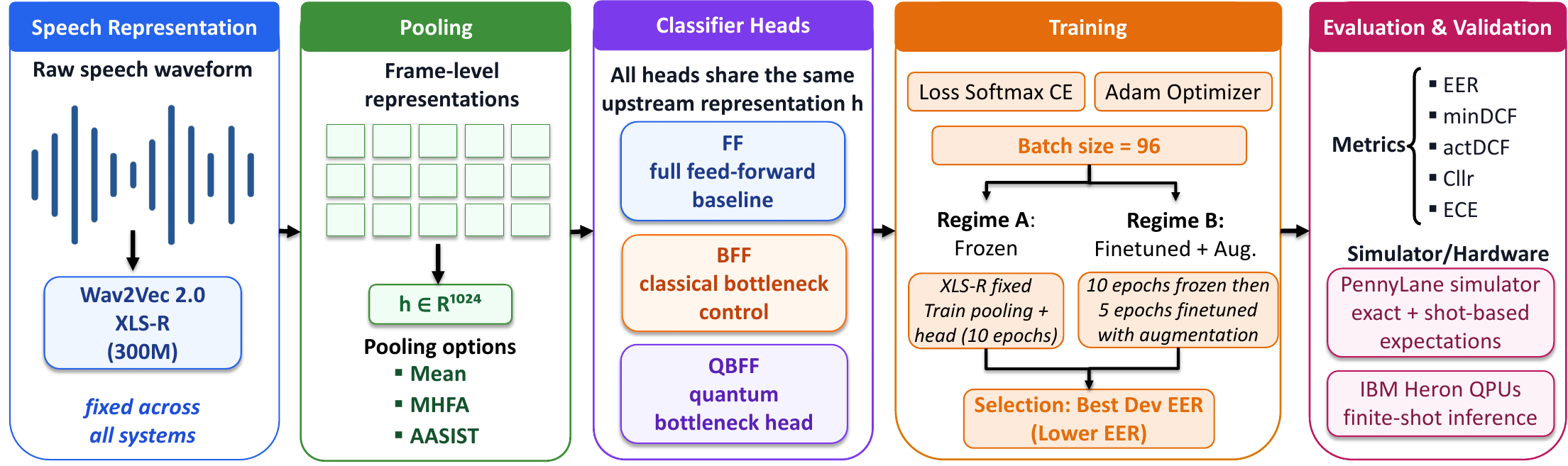}
    \caption{End-to-end experimental pipeline. A shared XLS-R front end and \texttt{Mean}, \texttt{MHFA}, or \texttt{AASIST} pooling produce a $1024$-dimensional utterance vector. It feeds either the full feed-forward baseline (\textit{FF}) or a $1024\rightarrow16\rightarrow8$ bottleneck head with a classical mixer (\textit{BFF}) or a variational quantum circuit (\textit{QBFF}). Systems are trained with a frozen front end or with fine-tuning and augmentation, then evaluated for discrimination, decision cost, and calibration in simulation; \textit{QBFF} is also tested on IBM Heron QPUs.}
    \label{method}
\end{figure*}
\subsection{Evaluation metrics}
\label{sec:metrics}

We use five metrics from the ASVspoof evaluation protocol~\cite{wang24_asvspoof}, grouped by what they capture.

\noindent\emph{Discrimination (threshold-free).}
\begin{itemize}[leftmargin=*]
\item \textbf{EER.}~\cite{BRUMMER2006230} Sweeping the threshold trades false accepts (spoof accepted as bona fide) against false rejects (bona fide rejected as spoof). At one threshold, the two rates are equal, and that shared rate is the EER. Lower is better: $0\%$ is perfect separation, $50\%$ is chance.
\item \textbf{Minimum detection cost (minDCF).} A cost-weighted error rate at its best achievable threshold. The cost model uses a bona fide prior $P_{\text{tar}}{=}0.95$ with error costs $C_{\text{miss}}{=}1$ for rejecting bona fide and $C_{\text{fa}}{=}10$ for accepting spoof, so accepting an attack is penalised ten times more than rejecting a genuine speaker. minDCF rewards a detector that \emph{can} be tuned to a low cost.
\end{itemize}

\noindent\emph{Deployment cost (at a fixed threshold).}
\begin{itemize}[leftmargin=*]
\item \textbf{actDCF.} The same cost-weighted error, evaluated at one fixed threshold rather than the best one. In our main tables, actDCF is computed from the pre-softmax log-likelihood-ratio (LLR) scores using the theoretical Bayes threshold defined by the stated ASVspoof 5 costs and priors, with no per-system tuning~\cite{wang24_asvspoof}. The gap between actDCF and minDCF quantifies the additional cost of using the fixed Bayes threshold instead of the empirically optimal one.
\item \textbf{Log-likelihood-ratio cost ($C_{llr}$).} An application-independent measure of score quality, in bits, that averages the detector's cost over the full range of operating points and penalizes confident wrong scores most. It is computed from the pre-softmax LLR scores and is not tied to a single threshold; a lower value means the scores remain useful across deployments.
\end{itemize}

\noindent\emph{Calibration.}
\begin{itemize}[leftmargin=*]
\item \textbf{Expected calibration error (ECE).} Group recordings into bins by predicted probability and, in each bin, compare the predicted probability to the observed fraction of bona fide. ECE is the average gap, weighted by bin size. Zero means the scores read exactly as probabilities.
\end{itemize}

\subsection{Quantum models for speech anti-spoofing}

Recent work on quantum speech anti-spoofing can be grouped by the role assigned to the quantum component. One research direction focuses on hybrid quantum-classical classifiers,  where acoustic features or learned speech representations are processed by parameterized quantum circuits, quantum neural layers, or quantum-enhanced CNN modules \cite{wang2023qtlwavlm,pandey2024deepfake,pandey2025hybridcqdeepfakeaudio,roy2026q}. These studies show that quantum circuits can be integrated into audio-forensics pipelines, but the reported gains are often confounded by changes in feature extraction, classifier design, training protocols, or model capacity. This makes it difficult to attribute performance differences to the quantum transformation itself.

Another direction explores quantum-kernel formulations for audio spoofing and deepfake detection. In these methods, acoustic descriptors, spectrogram features, or learned audio embeddings are mapped into a quantum feature space and compared through a quantum kernel or QSVM classifier \cite{alamins2025qksvm,amin2026quantum}. Kernel-based designs give a more precise representation-level interpretation, since the circuit defines the similarity measure used by the classifier. However, they usually focus on discrimination performance and are commonly evaluated using simulated kernels rather than as trainable quantum modules within a full self-supervised speech pipeline.

Related reliability research maps classical uncertainty-quantification methods to QML~\cite{wendlinger2025oldrules}, studies measurement-based uncertainty in physics-constrained learning~\cite{hewage2026measurement}, and evaluates VQCs using SAFE reliability criteria~\cite{chen2026safe}. QBFF complements these studies by evaluating score calibration and decision cost in a speech anti-spoofing pipeline with matched classical controls and real-QPU execution.

\begin{table}[t]
\centering
\caption{Quantum approaches to speech anti-spoofing, by capability and reported result. The highlighted columns indicate real-QPU execution and reliability evaluation. Accuracy shown where EER is not reported. ITW: In-the-Wild.}
\label{tab:quantum-comparison}
\footnotesize
\setlength{\tabcolsep}{2pt}
\begin{tabularx}{\linewidth}{@{}
>{\raggedright\arraybackslash}p{0.22\linewidth}
>{\raggedright\arraybackslash}p{0.19\linewidth}
>{\raggedright\arraybackslash}X
|>{\centering\arraybackslash}p{0.09\linewidth}
>{\centering\arraybackslash}p{0.12\linewidth}@{}}
\toprule
System & Quantum role & Benchmark (EER \%) &
\cellcolor{relcol}Real QPU &
\cellcolor{relcol}Reliab.\ eval \\
\midrule
Pandey et al.~\cite{pandey2025hybridcqdeepfakeaudio}
& Hybrid CNN--QNN
& Custom set (90.0\% acc.)
& \textcolor{black!45}{$\times$}
& \textcolor{black!45}{$\times$} \\

Al Amin et al.~\cite{alamins2025qksvm}
& Quantum kernel
& ASVspoof~5 (18.3)
& \textcolor{black!45}{$\times$}
& \textcolor{black!45}{$\times$} \\

Wang et al.~\cite{wang2023qtlwavlm}
& VQC classifier
& ASVspoof~21 DF (5.51)
& \textcolor{black!45}{$\times$}
& \textcolor{black!45}{$\times$} \\

\textbf{AASIST + QBFF (ours)}
& Quantum bottleneck
& \textbf{ASVspoof~5 (5.44), ITW (3.97)}
& \cg\checkmark
& \cg\checkmark \\
\bottomrule
\end{tabularx}
\end{table}

Our work differs from these approaches in both the role of the quantum circuit and the evaluation protocol (\autoref{tab:quantum-comparison}). The VQC is not an independent classifier and does not change the feature extractor or classifier capacity. Instead, it transforms $16$ encoded values into an $8$-dimensional representation inside a self-supervised speech pipeline. This comparison measures the effect of replacing the classical mixer
with the quantum module under matched architectural interfaces. Two further properties matter for near-term deployment: the trained circuit runs on real IBM Heron processors rather than in simulation, and the evaluation targets score reliability and deployment cost, not discrimination alone.

% The reported discrimination of prior quantum systems is also difficult to read as competitive in current terms. The quantum-kernel detector~\cite{alamins2025qksvm} reports $18.3\%$ EER on ASVspoof~5, far above current systems on that benchmark. The transfer-learning system~\cite{wang2023qtlwavlm} is evaluated on ASVspoof~2021, an older benchmark on which modern detectors now reach roughly $2$--$3\%$ EER~\cite{tak22_odyssey}, so its $5.51\%$ no longer tracks the state of the art. The hybrid classical-quantum model~\cite{pandey2025hybridcqdeepfakeaudio} reports accuracy on a custom recording set rather than on a community anti-spoofing benchmark, so its results do not transfer to standard evaluation. Our systems instead match current published results on ASVspoof~5 and the standard In-the-Wild set (\autoref{tab:sota-comparison}). The quantum bottleneck is therefore added to a detector that is already competitive with the state of the art, not to a weak baseline.

Prior quantum systems remain distant from strong classical baselines or use less comparable evaluations. The quantum-kernel detector reports $18.3\%$ EER on ASVspoof~5~\cite{alamins2025qksvm}; the transfer-learning system reports $5.51\%$ on ASVspoof~2021~\cite{wang2023qtlwavlm}, compared with roughly $2$--$3\%$ for strong classical detectors~\cite{tak22_odyssey}. The hybrid CNN--QNN reports accuracy on a custom dataset~\cite{pandey2025hybridcqdeepfakeaudio}. In contrast, \texttt{AASIST+QBFF} achieves $5.44\%$ EER on ASVspoof~5 and $3.97\%$ on In-the-Wild, making it competitive with several published classical systems (\autoref{tab:sota-comparison}). These cross-study comparisons provide context rather than controlled evidence of superiority.

% The closest prior work to our setup is~\cite{wang2023qtlwavlm}, which uses a VQC to transform features in an SSL-based pipeline. They report an ASVspoof21 eval EER of 5.51\%, improving their classical baseline from 6.8\%. In contrast, SOTA baselines already reach sub-3\% EER on ASVspoof21 eval~\cite{yamagishi21_asvspoof}, and our work evaluates score reliability rather than raw discrimination. More recently, \cite{alamins2025qksvm} proposes a quantum-kernel approach and reports an ASVspoof5 EER of 18.3\% for quantum-enabled models and 29\% for non-quantum, which deviates strongly from current SOTA baselines achieving sub-5\% EER on ASVspoof5 eval~\cite{wang24_asvspoof}.
%----------------------------------------
\section{Methodology}

\autoref{method} summarizes the pipeline. Within each pooling and training regime, we compare \textit{FF}, \textit{BFF}, and \textit{QBFF} using the same datasets and evaluation protocol.

\subsection{SSL feature extraction}

Let $x$ denote an input speech waveform. Each recording is truncated or zero-padded to a fixed duration of $4$ seconds at $16$ kHz, so all inputs share a common waveform length before feature extraction. The waveform is passed through a \texttt{Wav2Vec~2.0 XLS-R} 300M encoder~\cite{babu22_interspeech}, which maps the raw signal to a sequence of contextual frame-level speech representations,
\begin{equation}
\mathbf{H}=E_{\phi}(x)\in\mathbb{R}^{T\times d},
\end{equation}
where $T$ is the number of frames, $d=1024$ is the feature dimension, and $\phi$ denotes the encoder parameters. In the frozen regime, the encoder is fixed, so only the pooling module and classifier head are trained.

\subsection{Utterance pooling}

An utterance-level pooling module maps the frame representations to
\begin{equation}
\mathbf{h}=P_{\rho}(\mathbf{H})\in\mathbb{R}^{1024},
\end{equation}
where $\rho$ denotes trainable pooling parameters when applicable. We compare uniform temporal averaging (\texttt{Mean}) with learned attention-based pooling (\texttt{MHFA}~\cite{mhfa} and \texttt{AASIST}~\cite{aasist}) to assess dependence on the utterance representation.

\subsection{Classifier heads}
\begin{figure*}[htpb]
    \centering
    \includegraphics[width=1\linewidth]{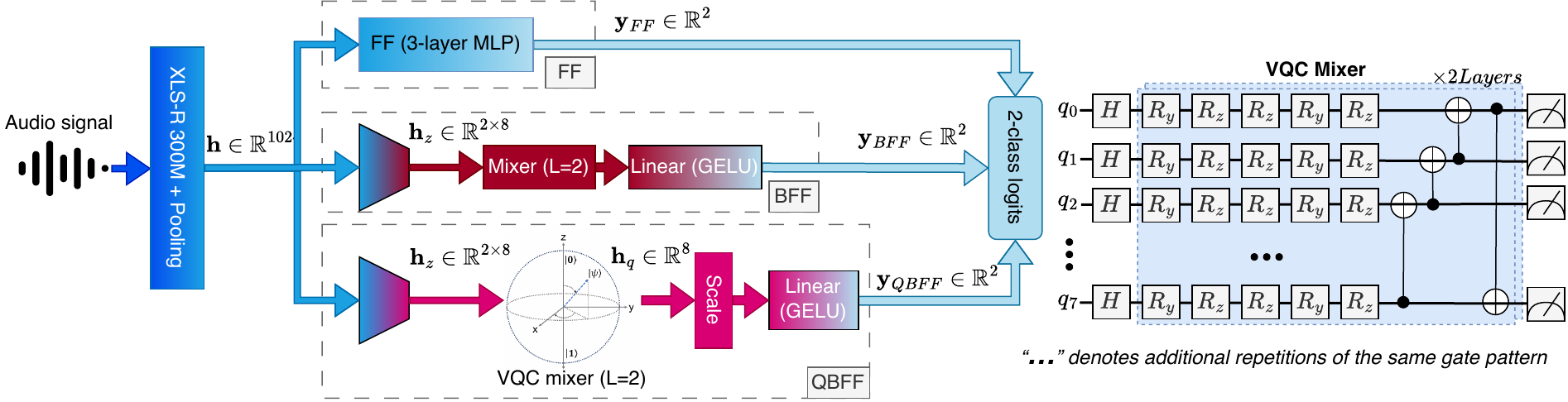}
    \caption{Classifier heads operating on pooled XLS-R embeddings. FF uses a $1024\rightarrow512\rightarrow256\rightarrow2$ MLP. BFF and QBFF project the embedding to $16$ bounded values and produce $8$ mixer outputs using a classical mixer or an $8$-qubit VQC, respectively. Both use an $8\rightarrow16\rightarrow2$ classifier; QBFF additionally learns an output scale.}
    \label{fig:architecture-overview}
\end{figure*}
The pooled embedding $\mathbf{h}$ is passed to one of three classifier heads, as shown in \autoref{fig:architecture-overview}. Each head maps the same utterance representation to two logits,
\begin{equation}
\boldsymbol{\ell}=f_{\theta}(\mathbf{h})\in\mathbb{R}^{2},
\end{equation}
where $\theta$ denotes the head parameters and the two logits correspond to bona fide and spoof classes. The heads differ in the transformation applied after pooling.

The feed-forward baseline \textit{FF} applies a three-layer MLP directly to the full $1024$-dimensional embedding,
\begin{equation}
f_{\mathrm{FF}}:\mathbb{R}^{1024}\rightarrow\mathbb{R}^{2},
\qquad
1024\rightarrow512\rightarrow256\rightarrow2.
\end{equation}
Each hidden layer is implemented as a Linear-BatchNorm-ReLU block, followed by a final linear layer that produces the two logits. This head provides the non-bottleneck reference.

Both bottleneck heads apply layer normalization, a linear projection from $1024$ to $16$ dimensions, and a second layer normalization:
\begin{equation}
\mathbf{z}=g_{\eta}(\mathbf{h})\in\mathbb{R}^{16}.
\end{equation}
Both then bound the projected values as
\begin{equation}
\boldsymbol{\alpha}=\pi\tanh(\mathbf{z})
\in(-\pi,\pi)^{16}.
\end{equation}
The classical head \textit{BFF} applies a two-layer mixer $m_{\beta}$ with dimensions $16\rightarrow8\rightarrow8$ and GELU activations:
\begin{equation}
\boldsymbol{\ell}_{\mathrm{BFF}}
=c_{\gamma}\!\left(m_{\beta}(\boldsymbol{\alpha})\right).
\end{equation}
The proposed \textit{QBFF} replaces this mixer with an $8$-qubit VQC receiving the same number of bounded input values. Both heads use a final classifier $c_{\gamma}$ with dimensions $8\rightarrow16\rightarrow2$, with GELU and dropout between its linear layers.

After Hadamard initialization, each of the $L=2$ circuit layers encodes $\alpha_j$ through $R_y$ and $\alpha_{j+8}$ through $R_z$ on qubit $j$, for $j=1,\ldots,8$. Each layer then applies three trainable rotations per qubit, in the order $R_z$, $R_y$, $R_z$, followed by a CNOT ring connecting $j$ to $j+1$ and qubit $8$ to qubit $1$. The same input angles are re-uploaded at each layer, giving $3\times8\times2=48$ trainable circuit parameters.

Let $U_{\omega}(\boldsymbol{\alpha})$ denote the full parameterized circuit, where $\omega$ contains the trainable quantum parameters. The state before measurement is
\begin{equation}
|\psi(\boldsymbol{\alpha};\omega)\rangle=U_{\omega}(\boldsymbol{\alpha})\mathsf{H}^{\otimes 8}|0\rangle^{\otimes 8}.
\end{equation}
The quantum bottleneck output is obtained by measuring Pauli-$Z$ expectation values on all qubits,
\begin{equation}
\mathbf{q}=Q_{\omega}(\boldsymbol{\alpha})=\left[\langle Z_1\rangle,\ldots,\langle Z_8\rangle\right]\in[-1,1]^8.
\end{equation}
The measurement outputs lie in $[-1,1]^8$, but a trainable scaling vector $\mathbf{a}\in\mathbb{R}^{8}$ rescales them before classification. Thus, bounded measurements do not by themselves guarantee calibrated predictions. The logits are
\begin{equation}
\boldsymbol{\ell}_{\mathrm{QBFF}}=c_{\gamma}\!\left(\mathbf{a}\odot\mathbf{q}\right),
\end{equation}
where $\odot$ denotes element-wise multiplication.

The comparison \textit{FF}$\rightarrow$\textit{BFF} evaluates the compact classical head, whereas \textit{BFF}$\rightarrow$\textit{QBFF} replaces the classical mixer with a VQC. The latter matches the projection and classifier architectures, mixer input and output dimensions, and nominal depth; each model is trained independently. The complete BFF and QBFF heads contain $18{,}866$ and $18{,}714$ trainable parameters, respectively, excluding pooling and the speech encoder.

\subsection{Training}

All systems are trained for binary anti-spoofing classification. Given logits $\boldsymbol{\ell}$ and label $y\in\{0,1\}$, the predicted posterior is
\begin{equation}
p(y\mid x)=\mathrm{softmax}(\boldsymbol{\ell})_y.
\end{equation}
For $N$ training examples $(x_i,y_i)$, where $x_i$ is a waveform and $y_i$ its class label, the objective is softmax cross-entropy,
\begin{equation}
\mathcal{L}=-\frac{1}{N}\sum_{i=1}^{N}\log p(y_i\mid x_i).
\end{equation}

We use two training regimes. In the \emph{frozen} regime, the \texttt{XLS-R} encoder remains fixed, and only the pooling module and classifier head are trained. This regime is used for the main ablation experiments because it provides a controlled setting in which changes in performance can be attributed to the pooling and head designs.

In the \emph{fine-tuned+augmentation} regime, training first uses the frozen setting, then fine-tunes the \texttt{XLS-R} encoder with audio augmentation. Following the SZU-AFS ASVspoof~5 recipe~\cite{xu24_asvspoof}, five augmentations expose the detector to common distortions and channel variation: silence trimming, time masking~\cite{specaugment}, Mu-law companding, RawBoost~\cite{rawboost}, and noise filtering. In all experiments, the checkpoint with the lowest development-set EER is selected. The optimizer, learning rate, batch size, epoch counts, and augmentation probabilities are reported in \autoref{sec:exp-setup}.

\subsection{Evaluation and Validation}

Each model outputs class posteriors using the softmax function. Unless stated otherwise, the spoof posterior is used as the detection score,
\begin{equation}
s(x)=p(\mathrm{spoof}\mid x).
\end{equation}

For actDCF and $C_{llr}$, we additionally retain and use the corresponding pre-softmax log-likelihood-ratio score. The same score definitions are used for \textit{FF}, \textit{BFF}, and \textit{QBFF}, allowing the heads to be compared under a common anti-spoofing protocol.

Evaluation uses both in-domain and out-of-domain test conditions, with the metrics defined in \autoref{sec}. Quantum models are evaluated using noiseless and finite-shot simulations, and the trained \textit{QBFF} circuit is then executed on IBM Heron-class processors to test its performance under finite sampling and hardware noise. Experimental settings and hardware-validation details are reported in \autoref{sec:exp-setup} and \autoref{sec:real_qpu}, respectively.
\section{Results and Analysis}
\label{sec:quant-bottleneck}

\subsection{Experimental Setup}
\label{sec:exp-setup}

All experiments are conducted on ASVspoof~5 Track~1~\cite{asvspoofdataset}, using the train partition for optimization, the development partition for checkpoint selection, and the evaluation partition for in-domain testing. Out-of-domain performance is evaluated on In-the-Wild~\cite{muller2022does}. The main simulation comparisons use three random seeds. Quantum simulation uses PennyLane~\cite{bergholm2018pennylane} and a native PyTorch state-vector implementation~\cite{firc2026vqcsim}. For real-device validation, the trained \textit{QBFF} circuit is executed through the Qiskit Runtime backend on the IBM QPU devices (Marrakesh and Torino) \cite{javadi2024quantum}. The experimental settings and augmentation probabilities are summarized in \autoref{tab:exp-setup}.

\begin{table}[htbp]
\centering
\caption{Experimental setup. Augmentations (right) apply only in the fine-tuned+augmentation regime, at the listed per-utterance probability~$p$.}
\label{tab:exp-setup}
\small
\setlength{\tabcolsep}{5pt}
\resizebox{0.9\columnwidth}{!}{%
\begin{tabular}{@{}ll@{\hspace{1.2em}}|@{\hspace{1.2em}}ll@{}}
\toprule
\textbf{Training setting} & \textbf{Value} & \textbf{Augmentation} & \textbf{$p$} \\
\midrule
Input length  & $4$ s, $16$ kHz   & Silence trimming  & $0.5$ \\
Optimizer     & Adam              & Time masking      & $0.3$ \\
Learning rate & $1\times10^{-4}$  & Mu-law companding & $0.3$ \\
Batch size    & $96$              & RawBoost          & $0.3$ \\
Frozen regime & $10$ epochs       & Noise filtering   & $0.3$ \\
Finetune+aug. & $10{+}5$ epochs   &                   &       \\
\bottomrule
\end{tabular}}
\end{table}

%%%%%%%%%%%%%%%%%%%%%
\begin{table*}[htpb]
\centering
\caption{Detection performance with a frozen encoder and no augmentation (mean$\pm$std over three seeds; lower is better). Embedding size denotes the pooled dimension for FF and the mixer output dimension for BFF/QBFF. Parameter counts cover the complete classifier head, excluding pooling and the encoder. Colors compare each row with the preceding head within its pooling block: \colorbox{goodcol}{better} or \colorbox{badcol}{worse}. Bold marks the best value within each block. Fine-tuned results appear in \autoref{tab:strong-metrics}.}
\label{tab:main_results}
\setlength{\tabcolsep}{3.2pt}
\renewcommand{\arraystretch}{1.10}
\resizebox{\linewidth}{!}{%
\begin{tabular}{llccccc|cccc|cc}
\toprule
& & & & & \multicolumn{4}{c}{\textbf{ASVspoof~5 Eval (In-Domain)}} & \multicolumn{4}{c}{\textbf{In-the-Wild (Out-of-Domain)}} \\
\cmidrule(lr){6-9} \cmidrule(lr){10-13}
& & & & & \multicolumn{2}{c}{Discrimination} & \multicolumn{2}{|c}{\cellcolor{relcol}\textbf{Reliability}} & \multicolumn{2}{c}{Discrimination} & \multicolumn{2}{|c}{\cellcolor{relcol}\textbf{Reliability}} \\
\cmidrule(lr){6-7} \cmidrule(lr){8-9} \cmidrule(lr){10-11} \cmidrule(lr){12-13}
\textbf{Pooling} & \textbf{Model} & \textbf{Emb. size} & \textbf{Depth} & \textbf{Params} &
\textbf{EER (\%)} $\downarrow$ & \textbf{minDCF} $\downarrow$ & \cellcolor{relcol}\textbf{actDCF} $\downarrow$ & \cellcolor{relcol}\textbf{$C_{llr}$} $\downarrow$ &
\textbf{EER (\%)} $\downarrow$ & \textbf{minDCF} $\downarrow$ & \cellcolor{relcol}\textbf{actDCF} $\downarrow$ & \cellcolor{relcol}\textbf{$C_{llr}$} $\downarrow$ \\
\midrule
\multirow{3}{*}{\textbf{\texttt{AASIST}}} & \textit{FF} (Baseline) & 1024 & 3 & $\sim$658k & 10.08\sd{0.64} & 0.285\sd{0.019} & 0.424\sd{0.110} & 1.249\sd{0.769} & 14.71\sd{3.86} & 0.370\sd{0.097} & 0.768\sd{0.165} & 3.076\sd{1.819} \\
 & \textit{BFF} (Classical) & 8 & 2 & $\sim$19k & \cg\textbf{9.46}\sd{0.80} & \cg\textbf{0.269}\sd{0.019} & \cg0.385\sd{0.087} & \cg0.736\sd{0.293} & \cg\textbf{10.96}\sd{1.09} & \cg\textbf{0.267}\sd{0.038} & \cg0.371\sd{0.184} & \cg\textbf{1.076}\sd{0.620} \\
 & \textit{QBFF} (Quantum) & 8 & 2 & $\sim$19k & \cb10.13\sd{0.48} & \cb0.287\sd{0.011} & \cg\textbf{0.377}\sd{0.136} & \cg\textbf{0.474}\sd{0.111} & \cb12.02\sd{3.08} & \cb0.290\sd{0.055} & \cg\textbf{0.363}\sd{0.068} & \cb1.129\sd{0.395} \\
\midrule
\multirow{3}{*}{\textbf{\texttt{MHFA}}} & \textit{FF} (Baseline) & 1024 & 3 & $\sim$658k & 8.05\sd{0.37} & 0.233\sd{0.011} & 0.278\sd{0.022} & 0.559\sd{0.108} & \textbf{8.26}\sd{1.06} & \textbf{0.229}\sd{0.031} & 0.659\sd{0.154} & 2.099\sd{0.795} \\
 & \textit{BFF} (Classical) & 8 & 2 & $\sim$19k & \cg7.06\sd{0.78} & \cg0.204\sd{0.022} & \cg0.228\sd{0.029} & \cg0.401\sd{0.185} & \cb9.06\sd{1.91} & \cb0.259\sd{0.056} & \cg0.413\sd{0.103} & \cg1.063\sd{0.236} \\
 & \textit{QBFF} (Quantum) & 8 & 2 & $\sim$19k & \cg\textbf{6.96}\sd{0.27} & \cg\textbf{0.201}\sd{0.007} & \cg\textbf{0.224}\sd{0.024} & \cg\textbf{0.209}\sd{0.098} & \cb11.98\sd{1.74} & \cb0.292\sd{0.049} & \cg\textbf{0.373}\sd{0.086} & \cg\textbf{0.359}\sd{0.615} \\
\midrule
\multirow{3}{*}{\textbf{\texttt{Mean}}} & \textit{FF} (Baseline) & 1024 & 3 & $\sim$658k & 6.42\sd{0.06} & 0.186\sd{0.001} & 0.263\sd{0.000} & 0.585\sd{0.009} & \textbf{13.54}\sd{6.01} & \textbf{0.377}\sd{0.161} & 0.690\sd{0.035} & 2.165\sd{0.405} \\
 & \textit{BFF} (Classical) & 8 & 2 & $\sim$19k & \cg6.20\sd{0.37} & \cg0.179\sd{0.011} & \cg\textbf{0.191}\sd{0.018} & \cg\textbf{0.283}\sd{0.019} & \cb15.47\sd{2.79} & \cb0.414\sd{0.080} & \cg\textbf{0.491}\sd{0.100} & \cg\textbf{1.112}\sd{0.381} \\
 & \textit{QBFF} (Quantum) & 8 & 2 & $\sim$19k & \cg\textbf{6.15}\sd{0.08} & \cg\textbf{0.178}\sd{0.003} & \cb0.217\sd{0.027} & \cb0.345\sd{0.135} & \cg14.47\sd{3.62} & \cg0.381\sd{0.103} & \cb0.607\sd{0.191} & \cb1.672\sd{0.990} \\
\bottomrule
\end{tabular}
}
\end{table*}
% We evaluate three pooling methods (\texttt{AASIST}, \texttt{MHFA}, \texttt{Mean}) with three classifier heads (\textit{FF}, \textit{BFF}, \textit{QBFF}), training each configuration with three random seeds (we report mean$\pm$std).
% \autoref{tab:main_results} summarizes discrimination (EER, minDCF) and deployment cost (actDCF, $C_{llr}$) on ASVspoof~5 Eval and In-the-Wild.
% We read the table through two matched comparisons: \textit{FF}$\rightarrow$\textit{BFF} isolates the bottleneck (compression and encoding), and \textit{BFF}$\rightarrow$\textit{QBFF} isolates the quantum circuit (classical vs.\ quantum mixer). Three findings recur. First, the bottleneck alone improves in-domain discrimination and steadies it under domain shift. Second, swapping in the quantum circuit leaves discrimination roughly unchanged but improves reliability at the fixed decision threshold, for the attentive poolings. Third, both effects survive a stronger training recipe and carry over to real quantum hardware. We take each in turn.

% We use the frozen-front-end, no-augmentation regime for all ablation suites because it provides a controlled setting in which changes in performance can be attributed to the pooling/head design rather than to representation drift from \texttt{XLS-R} finetuning or to augmentation choices. We report the finetuned+aug results as a separate robustness check to verify that the reliability trends persist under a stronger training recipe.

\subsection{Main Empirical Findings}

\autoref{tab:main_results} separates the effect of introducing a classical bottleneck (\textit{FF}$\rightarrow$\textit{BFF}) from replacing its mixer with a quantum circuit (\textit{BFF}$\rightarrow$\textit{QBFF}). The classical bottleneck substantially reduces classifier size while retaining competitive discrimination. Replacing its mixer with the VQC yields gains mainly in score reliability, but these depend on the pooling module and metric. Relative to \textit{BFF}, \textit{QBFF} lowers actDCF for \texttt{MHFA} and \texttt{AASIST} on both datasets, with particularly large reductions in $C_{llr}$ for \texttt{MHFA}. For \texttt{Mean} pooling, actDCF and $C_{llr}$ worsen on both datasets despite small EER improvements. These results motivate examining discrimination, calibration, and deployment cost separately.

\subsection{\texorpdfstring{Effect of Dimensionality Reduction (\textit{FF}$\rightarrow$\textit{BFF})}{Effect of Dimensionality Reduction (FF to BFF)}}
The bottleneck reduces classifier size while retaining competitive discrimination, consistent with a regularization effect~\cite{inf_bottleneck}.

In-domain, \textit{BFF} improves discrimination over \textit{FF} for the two attentive poolings (\autoref{tab:main_results}, left): \texttt{AASIST} from $10.08\%$ to $9.46\%$ EER, and \texttt{MHFA} from $8.05\%$ to $7.06\%$.

Out-of-domain, the benefit is model-dependent (\autoref{tab:main_results}, right). For \texttt{AASIST}, \textit{BFF} improves both mean EER and run-to-run stability ($14.71\%$ at $\sigma{=}3.86$ down to $10.96\%$ at $\sigma{=}1.09$). For \texttt{MHFA}, the changes in discrimination are mixed, so the generalization benefit is not universal.

\subsection{\texorpdfstring{Quantum Bottleneck Effects (\textit{BFF}$\rightarrow$\textit{QBFF})}{Quantum Bottleneck Effects (BFF to QBFF)}}
\label{sec:qkernel_effects}
Replacing the classical mixer (\textit{BFF}) with the VQC (\textit{QBFF}) lowers actDCF for \texttt{MHFA} and \texttt{AASIST} on both datasets, while discrimination results are mixed. This distinction motivates the calibration analysis in \autoref{sec:calibration}.

In-domain, the clearest case is \texttt{MHFA} on ASVspoof~5 Eval: \textit{QBFF} lowers actDCF ($0.228\rightarrow0.224$) and roughly halves $C_{llr}$ ($0.401\rightarrow0.209$) at similar EER ($7.06\%$ vs.\ $6.96\%$). Under domain shift it lowers actDCF for both \texttt{MHFA} ($0.413\rightarrow0.373$) and \texttt{AASIST} ($0.371\rightarrow0.363$), and for \texttt{MHFA} on In-the-Wild it cuts $C_{llr}$ sharply ($1.063\rightarrow0.359$); for \texttt{AASIST}, $C_{llr}$ stays comparable ($1.076$ vs.\ $1.129$). These results show improved decision cost despite mixed changes in discrimination.

For \texttt{Mean} pooling, \textit{QBFF} barely changes in-domain EER ($6.20\rightarrow6.15\%$) and worsens out-of-domain reliability (actDCF $0.491\rightarrow0.607$; $C_{llr}$ $1.112\rightarrow1.672$). One possible explanation is that learned pooling yields representations better suited to the quantum transformation than temporal averaging does, but the present experiments do not test this hypothesis.

\subsection{Validation under \texttt{XLS-R} Fine-tuning and Augmentations}
\label{ssec:stronger_baselines}

To address practical relevance, we also evaluate the advanced recipe %(\autoref{sec:train}) 
that uses augmentations and fine-tunes the XLS-R. This yields SOTA-competitive discrimination for classical heads: \textit{AASIST+FF} reaches 4.71\% EER on ASVspoof~5 Eval, placing our baselines within the top range of published systems~\cite{wang24_asvspoof}.
For the two attentive poolings, the reliability gains of \textit{QBFF} over the matched \textit{BFF} persist under this stronger recipe, and on \texttt{AASIST} they strengthen across every metric. \texttt{Mean} stays the exception seen in the frozen regime. In-the-Wild results for all three poolings are in \autoref{tab:strong-metrics}. 
% The same class-stratified bootstrap confirms that the actDCF reduction remains significant ($p<0.001$), with $C_{llr}$ following suit.

\begin{table}[t]
\centering
\small
\caption{In-the-Wild performance under the strong fine-tuned+augmentation recipe (mean$\pm$std over 3 seeds). Color tint, \textbf{bold}, and the capped \emph{Reliability} columns follow \autoref{tab:main_results}.}
\label{tab:strong-metrics}
\resizebox{0.85\linewidth}{!}{
\begin{tabular}{lc|cc}
\toprule
& & \multicolumn{2}{|c}{\cellcolor{relcol}\textbf{Reliability}} \\
\cmidrule(lr){3-4}
\textbf{System} & \textbf{EER (\%)} $\downarrow$ & \cellcolor{relcol}\textbf{actDCF} $\downarrow$ & \cellcolor{relcol}\textbf{$C_{llr}$} $\downarrow$ \\
\midrule
\textit{AASIST + FF}   & 4.74\sd{0.76} & 0.286\sd{0.017} & 1.495\sd{0.064} \\
\textit{AASIST + BFF}  & \cb5.49\sd{0.23} & \cg0.205\sd{0.046} & \cg1.010\sd{0.283} \\
\textit{AASIST + QBFF} & \cg\textbf{3.97}\sd{0.10} & \cg\textbf{0.181}\sd{0.078} & \cg\textbf{0.934}\sd{0.503} \\
\midrule
\textit{MHFA + FF}   & \textbf{4.98}\sd{0.07} & 0.350\sd{0.119} & 1.631\sd{0.834} \\
\textit{MHFA + BFF}  & \cb6.25\sd{0.50} & \cg0.305\sd{0.048} & \cg1.465\sd{0.178} \\
\textit{MHFA + QBFF} & \cb6.61\sd{0.64} & \cg\textbf{0.262}\sd{0.024} & \cg\textbf{1.132}\sd{0.147} \\
\midrule
\textit{Mean + FF}   & \textbf{4.29}\sd{0.30} & 0.496\sd{0.101} & 2.476\sd{0.557} \\
\textit{Mean + BFF}  & \cb5.15\sd{0.56} & \cg\textbf{0.253}\sd{0.036} & \cg1.296\sd{0.246} \\
\textit{Mean + QBFF} & \cb8.98\sd{3.26} & \cb0.272\sd{0.096} & \cg\textbf{1.129}\sd{0.522} \\
\bottomrule
\end{tabular}}
\end{table}

The benefit, therefore, holds even on strong, fully tuned backbones, not only in the controlled frozen setting.

\subsection{Comparison with Published Systems}
\label{ssec:sota-comparison}

Our fully tuned systems achieve discrimination performance competitive with the published systems listed in \autoref{tab:sota-comparison}. On In-the-Wild, \textit{AASIST+QBFF} achieves $3.97\%$ EER, compared with $4.74\%$ for \textit{AASIST+FF}. ResNet-101$_{\mathrm{SV}}$ reports a lower EER of $2.16\%$ but uses additional external training datasets. On ASVspoof~5 Eval, \textit{AASIST+QBFF} instead has a higher EER than \textit{AASIST+FF} ($5.44\%$ versus $4.71\%$), showing that its discrimination benefit is dataset-dependent.

Because \texttt{XLS-R} is pretrained on external speech, the ASVspoof~5 Eval results provide context rather than a challenge-compliant ranking. We therefore emphasize the unrestricted In-the-Wild comparison and assess score reliability separately in \autoref{sec:calibration}.

\begin{table}[t]
\centering
\caption{Comparison with published anti-spoofing systems. The capped In-the-Wild column carries the comparison; ASVspoof~5 Eval is indicative only (see text). \textit{AASIST}~\cite{tak22_odyssey} is the same architecture as our \textit{AASIST+FF}, to confirm our reproduction.}
\label{tab:sota-comparison}
\resizebox{0.95\linewidth}{!}{
\begin{tabular}{lcc}
\toprule
System & ASVspoof 5 EER (\%) $\downarrow$ & \cellcolor{relcol}\textbf{In-the-Wild EER (\%)} $\downarrow$ \\
\midrule
\multicolumn{3}{@{}l}{\textit{Published systems}} \\
AASIST~\cite{tak22_odyssey} & 4.71 & 4.74 \\
ResNet-101$_{\text{SV}}$~\cite{dao24_asvspoof}\textsuperscript{$\dagger$} & \textit{n/a} & 2.16 \\
SLIM~\cite{zhu24_asvspoof} & 5.5 & 10.80 \\
P-V AASIST~\cite{guo24_asvspoof} & 5.99 & 6.05 \\
\midrule
\multicolumn{3}{@{}l}{\textit{This work}} \\
AASIST + FF   & 4.71 & 4.74 \\
AASIST + QBFF & 5.44 & \textbf{3.97} \\
MHFA + FF     & 6.05 & 4.98 \\
MHFA + QBFF   & 6.25 & 6.61 \\
\bottomrule
\end{tabular}}
\vspace{1pt}

{\scriptsize \textsuperscript{$\dagger$} ResNet-101$_{\text{SV}}$ uses additional external datasets for training~\cite{dao24_asvspoof}.}
\end{table}

\subsection{Calibration and Deployment Risk}
\label{sec:calibration}

\paragraph{Motivation}
Detectors are typically calibrated before deployment~\cite{wang24_asvspoof}, and miscalibration can introduce thresholded errors even when discrimination remains unchanged.
We therefore evaluate both intrinsic probability reliability and Dev-tuned deployment performance on Eval.

% \paragraph{Metrics.}
% We measure calibration with Expected Calibration Error (ECE) and report deployment risk with $actDCF_{\mathrm{dev}}$, i.e., $actDCF$ on Eval using a threshold selected on Dev.
% Lower values indicate more reliable posteriors and lower operating-point risk, respectively.

\begin{figure}[t]
    \centering
    \includegraphics[width=0.8\linewidth]{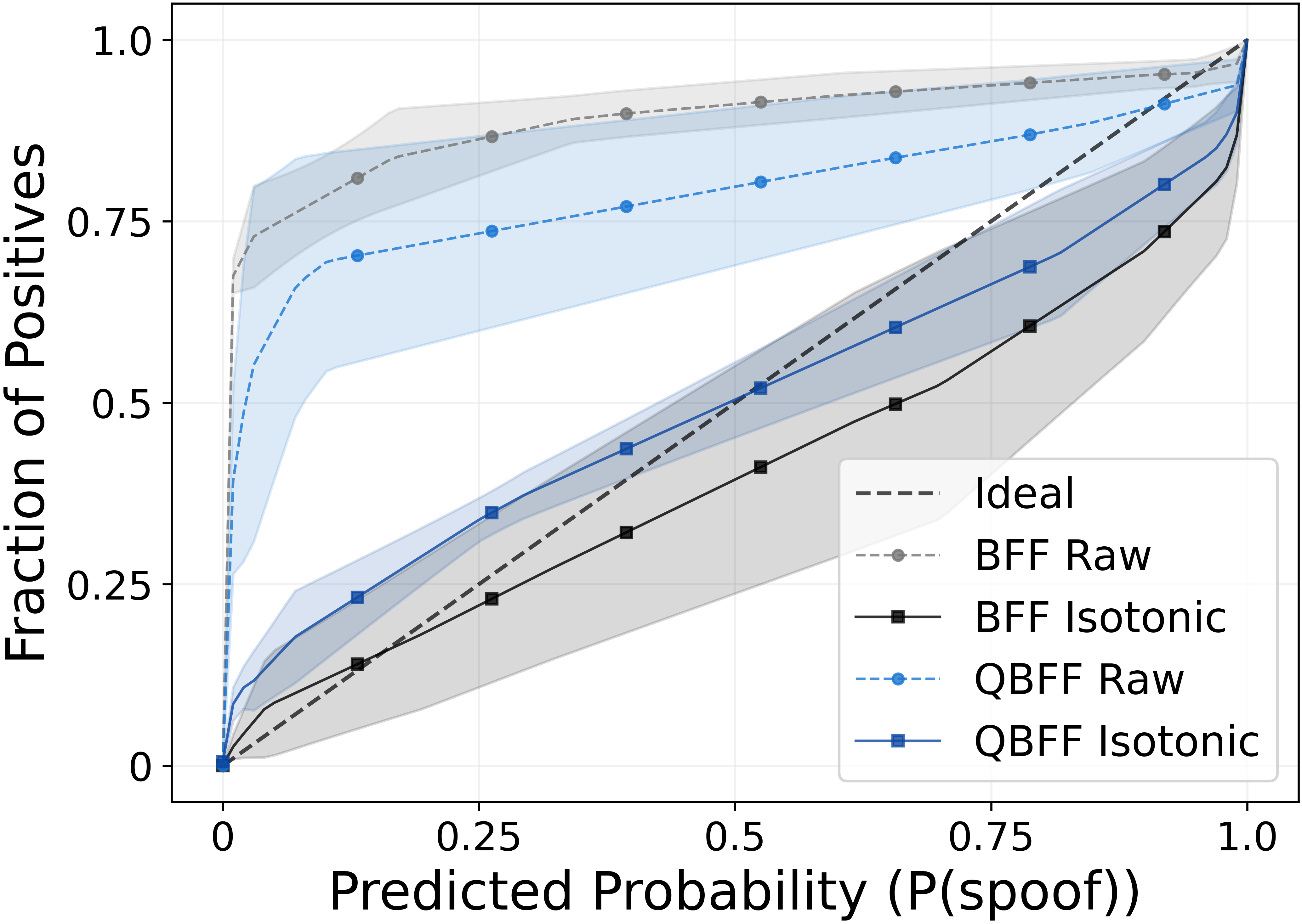}
    \caption{In-the-Wild reliability diagrams for MHFA heads before and after isotonic calibration fitted on ASVspoof~5 Dev. Curves average three seeds after interpolation of ten quantile-bin estimates to a common grid; shading shows $\pm$ one standard deviation. The diagonal indicates perfect calibration. In-domain results appear in \autoref{tab:calibration}.}
    \label{fig:calibration_panels}
\end{figure}

\paragraph{Protocol (tune on Dev, deploy on Eval)}
All systems output posteriors, converted to $P(\mathrm{spoof}\mid x)$.
We evaluate three score variants: uncalibrated (\textit{Raw}), Platt scaling, and isotonic regression.
Each calibrator is fitted on ASVspoof~5 Dev scores and applied to Eval scores.
For \autoref{tab:calibration}, ECE uses $15$ equal-width probability bins, weighted by their sample counts.
For each score stream, we select the DCF-optimal threshold on Dev and apply it on Eval, yielding $actDCF_{\mathrm{dev}}$ (\autoref{tab:calibration}, \autoref{fig:calibration_panels}).

\paragraph{Results}
For both poolings in \autoref{tab:calibration}, \textit{QBFF} has lower raw ECE than \textit{BFF}: $0.169\rightarrow0.078$ for \texttt{MHFA} and $0.190\rightarrow0.095$ for \texttt{AASIST}. After post-hoc calibration, however, \textit{BFF} achieves a lower best ECE for both poolings. Deployment cost shows a different pattern: the best $actDCF_{\mathrm{dev}}$ is lower for \textit{QBFF} than \textit{BFF} with \texttt{MHFA} ($0.221$ versus $0.230$), but higher with \texttt{AASIST} ($0.306$ versus $0.278$). These results show improved raw calibration with both poolings, with an additional reduction in Dev-tuned deployment cost for \texttt{MHFA}.

\begin{table}[t]
\centering
\caption{In-domain calibration and deployment risk on ASVspoof~5 Eval (mean over three seeds; out-of-domain reliability in \autoref{fig:calibration_panels}). \textit{Raw} = network posteriors as $P(\mathrm{spoof}\mid x)$; \textit{Best} = the post-hoc calibrator minimizing that metric; $actDCF_{\mathrm{dev}}$ uses a Dev-tuned threshold. Color tint as in \autoref{tab:main_results}.}
\label{tab:calibration}
\resizebox{\linewidth}{!}{%
\begin{tabular}{lcccc}
\toprule
System & Raw ECE $\downarrow$ & Best ECE $\downarrow$ &
$actDCF_{\mathrm{dev}}$ (Raw) $\downarrow$ & Best $actDCF_{\mathrm{dev}}$ $\downarrow$ \\
\midrule
\textit{MHFA + FF}   & 0.198 & 0.025 & 0.256 & 0.251 \\
\textit{MHFA + BFF}  & \cg0.169 & \cg0.023 & \cg0.234 & \cg0.230 \\
\textit{MHFA + QBFF} & \cg0.078 & \cb0.028 & \cg0.223 & \cg0.221 \\
\midrule
\textit{AASIST + FF}   & 0.348 & 0.054 & 0.291 & 0.287 \\
\textit{AASIST + BFF}  & \cg0.190 & \cg0.039 & \cg0.279 & \cg0.278 \\
\textit{AASIST + QBFF} & \cg0.095 & \cb0.051 & \cb0.306 & \cb0.306 \\
\bottomrule
\end{tabular}}
\end{table}

\subsection{Circuit Parameter Perturbation}
\label{ssec:checks}
We assess sensitivity to the trained circuit parameters by replacing them at inference time while keeping all other parameters, including the output scale, fixed. We use independent replacements drawn from $\mathcal{N}(0,0.1^2)$ or $\mathcal{U}(0,2\pi)$. Uniform replacement increases actDCF, showing that the detector's score reliability is sensitive to the trained circuit parameters.

\begin{table}[t]
\centering
\caption{Real-hardware evaluation of MHFA+QBFF on a hardness-stratified ASVspoof~5 subset. \textit{Mit.}~= reported mitigation grouping (DD+GM); \checkmark~= labeled full mitigation. Cell color compares each hardware run against the \textit{Classical} reference row (\colorbox{goodcol}{better}/\colorbox{badcol}{worse}); the capped \emph{Reliability} columns and \textbf{bold} best follow \autoref{tab:main_results}.}
\label{tab:real-hardware}
\resizebox{\linewidth}{!}{
\begin{tabular}{lcrcr|cc}
\toprule
Shots & Mit. & QPU(s) & \#Runs & EER (\%) $\downarrow$ & \cellcolor{relcol}\textbf{actDCF} $\downarrow$ & \cellcolor{relcol}\textbf{$C_{llr}$} $\downarrow$ \\
\midrule
-- & -- & Classical & 3 & 7.29 & 0.227 & 0.523 \\
\midrule
256 & varies & Marrakesh & 12 & \cg6.4\sd{2.1} & \cb0.25\sd{0.08} & \cg0.29\sd{0.14} \\
256 & \checkmark & Marrakesh & 4 & \cg5.9\sd{1.8} & \cg0.21\sd{0.07} & \cg0.21\sd{0.04} \\
512 & -- & Marrakesh+Torino & 5 & \cg4.7\sd{1.6} & \cb0.62\sd{0.65} & \cg0.26\sd{0.21} \\
512 & \checkmark & Marrakesh+Torino & 3 & \cg\textbf{4.69}\sd{1.56} & \cg\textbf{0.144}\sd{0.016} & \cg\textbf{0.156}\sd{0.065} \\
1024 & -- & Marrakesh & 1 & \cg6.25 & \cb1.841 & \cb0.610 \\
\bottomrule
\end{tabular}}
\end{table}

\subsection{Real Quantum Hardware (Feasibility)}
\label{sec:real_qpu}

We execute the trained \texttt{MHFA+QBFF} circuit on IBM Heron-class processors (Marrakesh and Torino)~\cite{javadi2024quantum}. \autoref{tab:real-hardware} reports runs at $256$, $512$, and $1024$ shots. The five-run $512$-shot row covers all recorded executions on the seed-$0$ 128-trial protocol; the three-run row follows the reported mitigation grouping. Relative to the \textit{Classical} row, the five-run row has lower mean EER and $C_{llr}$, but higher mean actDCF and substantial run-to-run variation. These results establish hardware feasibility without demonstrating a consistent deployment-cost improvement.
Because the mitigation settings associated with these QPU runs were not independently verified, the observed differences between mitigation groups cannot be attributed specifically to mitigation.
%Because the submitted QPU options have not been independently verified, differences between the reported mitigation groups cannot be attributed to mitigation.

The subset generator ranks a reference BFF model's class-oriented decision margins separately within each class and divides them into hard, medium, and easy thirds. Evaluation on this selected subset is not directly comparable to the full benchmark tables. The released repository includes the generator, trial protocols, and per-trial margin metadata.

For hardware evaluation, the saved score is $p=P(\mathrm{bona\ fide}\mid x)$. We compute actDCF using posterior log odds $\log[p/(1-p)]$ with threshold $\log 10$ and the stated error costs and prior. We evaluate the $C_{llr}$ functional on these scores as the class-averaged binary log loss in bits; the legacy hardware logs use twice this normalization and are converted accordingly.

\section{Conclusions}
\label{sec:conclusion}

We introduced \textit{QBFF}, an 8-qubit variational quantum bottleneck for self-supervised speech anti-spoofing. Compared with the matched BFF control, QBFF maintains broadly comparable discrimination while improving score calibration and deployment-oriented reliability across several attentive-pooling settings, including out-of-domain evaluation. Real-QPU execution establishes feasibility, but lower subset EER and $C_{llr}$ coexist with higher mean actDCF and substantial hardware-run variability. The benefits depend on the pooling and training regime: QBFF does not consistently improve Mean-pooling performance, and its lower raw ECE for AASIST does not translate into lower Dev-tuned actDCF. A bounded-output classical control is needed to isolate the role of measurement bounds. Our results support quantum bottlenecks as hardware-executable components for improving score reliability, evaluated through matched controls and calibration metrics alongside discrimination performance.

\begin{comment}
\subsection{Limitations}
\label{ssec:limitations}
These results are a proof-of-concept rather than a claim of universal quantum advantage.
Our real-QPU results use a fixed compact subset and therefore do not benchmark full ASVspoof~5-scale performance.
All experiments use an XLSR front-end, so generalization to other encoders is not tested here.
We do not fine-tune the XLSR extractor, nor do we use augmentations during training.
On the quantum side, we evaluate a single VQC configuration, leaving alternative ans\"atze unexplored.
Finally, reliability depends on finite-shot inference, and 512 shots is the most consistent regime in our runs.    
\end{comment}

% --------------------------------------------------------
\section*{Acknowledgment}
This work was supported by the Brno University of Technology internal project FIT-S-26-9011. Computational resources were provided by the e-INFRA CZ project (ID:90254), supported by the Ministry of Education, Youth and Sports of the Czech Republic. This work was also supported in part by the NYUAD Center for Quantum and Topological Systems (CQTS), funded by Tamkeen under the NYUAD Research Institute grant CG008, and the Center for Cyber Security (CCS), funded by Tamkeen under the NYUAD Research Institute Award G1104. The authors used ChatGPT and Claude for copy-editing, reviewed all AI-assisted text, and take full responsibility for the content.

\bibliographystyle{IEEEtran}
\bibliography{mybib}

\end{document}

%% file: img/abstract_fig.tex
\begin{figure}[t]
\centering
\resizebox{\columnwidth}{!}{%
\begin{tikzpicture}[
x=1mm,y=1mm,font=\sffamily\footnotesize,
box/.style={rounded corners=1.7pt,draw=#1!80!black,fill=#1!10,
line width=0.75pt,align=center,inner xsep=2.8pt,inner ysep=2.5pt,
minimum height=8mm},
input/.style={box=#1,text width=20mm},
bottle/.style={box=#1,text width=16mm},
pathbox/.style={box=#1,text width=25mm,minimum height=11mm},
score/.style={box=#1,text width=23mm},
hw/.style={box=#1,text width=28mm},
arrow/.style={-{Latex[length=2mm,width=1.4mm]},thick,draw=black!65},
softarrow/.style={-{Latex[length=1.8mm,width=1.2mm]},thick,draw=black!50},
hwarrow/.style={-{Latex[length=1.8mm,width=1.2mm]},thick,dashed,draw=purple!75!black}
]
\draw[rounded corners=2.5pt,dashed,draw=black!38,line width=0.55pt]
    (39,-19) rectangle (105,19);
\node[font=\sffamily\scriptsize,fill=white,inner xsep=1.5pt,anchor=west]
    at (40,19) {\textbf{C2: matched comparison}};
\node[input=blue] (front) at (0,0)
    {\textbf{Speech}\\1024-D embedding};
\node[bottle=gray] (bneck) at (26,0)
    {\textbf{Projection}\\$1024\to16$\\bounded inputs};
\node[pathbox=purple] (qbff) at (55,9)
    {\textbf{C1: QBFF}\\8-qubit VQC\\$Q_\omega(\boldsymbol{\alpha})\in\mathbb{R}^{8}$};
\node[pathbox=orange] (bff) at (55,-9)
    {\textbf{BFF control}\\classical mixer\\$m_\beta(\boldsymbol{\alpha})\in\mathbb{R}^{8}$};
\node[score=magenta] (score) at (89,0)
    {\textbf{Matched classifier}\\$8\to16\to2$};
\node[hw=purple] (hardware) at (91,30)
    {\textbf{C3: QPU run}\\IBM Heron\\finite shots};
\draw[arrow] (front) -- (bneck);
\draw[softarrow] (bneck.east) -- (qbff.west);
\draw[softarrow] (bneck.east) -- (bff.west);
\draw[softarrow] (qbff.east) -- (score.west);
\draw[softarrow] (bff.east) -- (score.west);
\draw[hwarrow] (qbff.north east) -- (hardware.south west);
\end{tikzpicture}%
}
\caption{QBFF and BFF use matched projection and classifier architectures, trained independently. Their mixers map $16$ bounded inputs to $8$ outputs; QBFF additionally applies a learned output scale (omitted). The comparison evaluates discrimination and score reliability, with real-QPU execution assessing feasibility.}
\label{fig:contribution}
\end{figure}